\documentclass[prd,a4paper]{revtex4}

\usepackage[pdftex]{graphicx}
\usepackage{inputenc}
\usepackage{amsmath}
\usepackage{amssymb}
\usepackage{subfigure}
\usepackage{hyperref}

\inputencoding{latin9}

\newcommand{\nc}{\newcommand}

\def\lsim{\; \raise0.3ex\hbox{$<$\kern-0.75em
      \raise-1.1ex\hbox{$\sim$}}\; }
\def\gsim{\; \raise0.3ex\hbox{$>$\kern-0.75em
      \raise-1.1\textmd{}ex\hbox{$\sim$}}\; }

\nc{\be}[1]{\begin{equation}\mbox{$\label{#1}$}}
\nc{\bea}[1]{\begin{eqnarray} \mbox{$\label{#1}$}}
\nc{\Section}[2]{\section{#2}\label{#1}}
\nc{\Bibitem}[1]{\bibitem{#1}}
\nc{\Label}[1]{\label{#1}}
\nc{\ie}{{\em i.e. }}
\nc{\eg}{{\em e.g. }}
\nc{\eea}{\end{eqnarray}}
\nc{\ee}{\end{equation}}
\nc{\w}{\omega}
\begin{document}

\title{CUDAEASY - a GPU Accelerated Cosmological Lattice Program}

\author{J. Sainio}
\thanks{jani.sainio@utu.fi}
\affiliation{Department of Physics and Astronomy,
University of Turku, FIN-20014 Turku, FINLAND}
%\author{I. Vilja}
%\thanks{vilja@utu.fi}
%\affiliation{Department of Physics,
%University of Turku, FIN-20014 Turku, FINLAND}
\date{\today}

\begin{abstract}

This paper presents, to the author's knowledge, the first graphics processing unit (GPU) accelerated program
that solves the evolution of interacting scalar fields in an expanding universe.
We present the implementation in NVIDIA's Compute Unified Device Architecture (CUDA) and
compare the performance to other similar programs in chaotic inflation models.
We report speedups between one and two orders of magnitude depending on the used hardware and software
while achieving small errors in single precision.
Simulations that used to last roughly one day to compute can now be done in hours and this difference is expected to increase in the future.
The program has been written in the spirit of LATTICEEASY and users of the aforementioned program should find it relatively easy to start using CUDAEASY in lattice simulations. The program is available at \href{http://www.physics.utu.fi/theory/particlecosmology/cudaeasy/}{http://www.physics.utu.fi/theory/particlecosmology/cudaeasy/} under
the GNU General Public License.

\end{abstract}

\maketitle

\section{Introduction}

%Recent developments in computer technology have shifted the emphasis from fast single core processors into multicore processors.
%This parallelization naturally needs a different line of thought compared to the old single core
%architecture in order to leverage the available computational potential. 

Recent developments in computer technology have made graphic processing units (GPUs) available to scientists as a powerful coprocessor in numerical computations. For example the theoretical peak performance of a state of the art AMD 5870 GPU is 2.7 tera floating point operations per second (TFLOPS) in single precision \cite{ATI:2009} compared to the peak performance 55.36 giga flops (GFLOPS) of Intel's Core i7 \cite{Intel:2009}. This difference in computational horsepower has made general-purpose computing on graphics processing units (GPGPU) an increasingly popular way to do numerical computations \cite{Belleman:2007kv,Brunner:2007sy,Zwart:2008bp,Ishikawa:2008pf,Szalay:2008,Anselmi:2008hi,Ford:2008em,Ord:2009xk,Gaburov:2009dy,Demchik:2009ni,Chung:2009yb,Schive:2009hw,Jonsson:2009dh,Groen:2009,Nakasato:2009xq,Khanna:2009zs,Hagiwara:2009cy,CapuzzoDolcetta:2009er,Banerjee:2009hs,Wang:2009st}. 

Due to the inherently parallel nature of computer graphics GPUs are well suited for problems that are easy to parallelize.
In ideal situations a GPU version of a program can be up to three orders of magnitude faster \cite{Januszewski:2009} than the serial CPU version of the same program. Speedups of this magnitude can have dramatic effects on how science is made and make some previously computationally difficult problems rather trivial.

The shear computational power of GPUs is however useless if if cannot be harnessed effectively. Therefore the use of computationally suitable language is of paramount importance. Recent developments in this sector have made programming of modern GPUs relatively easy. NVIDIA's CUDA architecture \cite{NVIDIA:2009} and the OpenCL language \cite{OpenCL:2009} of the Khronos Group are similar to C syntax and therefore lower considerably the learning curve needed to write programs that use GPUs as co-processors. How GPUs are programmed in these languages is fairly similar and by learning one of these makes the other easy to learn as well.

Cosmology has many computationally demanding problems \cite{Hoerner:1960,Hoerner:1963,Bertschinger:1998}. The study of interacting fields and reheating after inflation in early universe is one of them. %the most studied computational problem in cosmology.
The nonlinear nature of the system compels to study the evolution of the system numerically.
The problem then becomes the distretization of the scalar field equations and solving the evolution of the system in a lattice once the initial values have been set.
This problem has been previously solved in LATTICEEASY \cite{Felder:2000hq} and DEFROST \cite{Frolov:2008hy} programs that use very different methods to solve the dynamics of the system.

The lattice calculations are relatively easily parallelizable and therefore suitable for GPU computing. The aim of this paper is to present a GPU implementation of a program that solves the evolution of interacting scalar fields in an expanding universe. The main evolution algorithm is identical to the leap-frog method of LATTICEEASY but as was pointed out in \cite{Frolov:2008hy} the initialization of LATTICEEASY is not entirely consistent. We have therefore adapted the initialization method from DEFROST and implemented this in the GPU. Our program can be therefore thought as an amalgam of the LATTICEEASY and DEFROST codes.

We have chosen to write the program in the CUDA architecture mainly for pragmatig reasons: when we started to develop this code OpenCL was still in its infancy whereas CUDA was many generations old and had comprehensive documentation and a large group of developers. Note however that since OpenCL code can be run in both AMD and NVIDIA GPUs the future \textsl{de facto} GPU programming language is still undecided. In order to future proof the CUDAEASY program we are looking into porting it also into the OpenCL language.

This paper is organized as follows: in section II we will present the CUDA architecture, the equations of motion and our implementation of the lattice program. We will present numerical results in section III and we will conclude in section IV.

\section{GPU implementation}

\subsection{CUDA programming model}

NVIDIA introduced CUDA (an acronym for Compute Unified Device Architecture) in November 2006 as a programming language for NVIDIA GPUs \cite{NVIDIA:2009}. CUDA comes with C for CUDA application programming interface that gives programmers familiar with C an easy way to start writing programs that are executed on GPU. This is acchieved by extending C with different CUDA commands. 

Within the CUDA framework GPUs are called \textsl{devices} whereas the CPU is known as the \textsl{host}. The fundamental concept in GPGPU are the lightweight \textsl{threads} that are executed in parallel on the device with \textsl{kernel} functions. Threads within a kernel constitute a two dimensional \textsl{grid} that is divided into different \textsl{blocks} that hold the threads. Blocks can be one, two or three dimensional and threads within the same block can share information through a \textsl{shared memory}. Threads and blocks have ids (threadIdx and blockIdx respectively) which can be used to calculate the global thread indexes within the grid.

The device has a range of different memories that have different characteristics and roles in computing. \textsl{Global} or \textsl{device memory} is the main memory of the GPU but it is located off-chip and therefore has a considerable latency. Reads from the global memory should also be \textsl{coalesced}: threads within a block are divided into \textsl{warps} that are made of 32 threads. Threads within a \textsl{half-warp} (\ie 16 threads) should read from a same segment of memory which size depends on the type of data. For example for floats this should be 128 bytes. If this requirement is not met there will be a performance penalty. The GPU also has a limited amount of fast shared memory that can be used to hide the latency of the global memory and to share data between threads within a block in order to save the memory bandwidth. The device also has \textsl{registers} and read-only \textsl{texture} and \textsl{constant memories}. A more accurate description of these can be found in the CUDA programming guide. We will be using a range of these memories in order to optimize the execution of the kernel.

The device in NVIDIA GT200 series is divided into \textsl{streaming multiprocessors} (SM) that do the actual computations. Each SM has 8 single precision calculating units, 1 double precision unit and 2 super function units. Shared memory and registers are also located on the SMs. Once a kernel is launched on the device the resource usage of the kernel dictates how many blocks a SM can execute. Therefore it is crucial to keep the register and shared memory usage of a thread to its minimum whilst also minimizing the global memory fetches. The global reads should also be coalesced for the performance to be optimal which means that number of threads block should be a multiple of 16. The search for the optimal number of threads per block is however largely an exercise in trial and error.

\subsection{Equations of motion}

The dynamics of early universe are often modelled with classical scalar fields that are minimally coupled to gravity and interact with each other through the potential term. Starting from the action principle and by demanding Lorentz invariance the action is compelled to be a function of only the fields and their temporal and spatial first derivatives. Variation of this action with respect to the fields naturally leads to second order differential equations. In the case of reheating there are $N$ scalar fields $\phi_{i}, \, i=1,...,N$ which interact through potential $V(\phi_{1},...,\phi_{N})$ which leads to equations of motion
\begin{equation} \label{eq:fe}
\ddot{\phi_{i}}+3H\dot{\phi_{i}}-\frac{\Delta}{a^2}\phi_{i}+\frac{\partial V}{\partial \phi_{i}} = 0,
\end{equation}
where $H=\dot{a}/a$, $\Delta$ is the Laplacian with respect to comoving coordinates and dot denotes differentiation with respect to physical time.

The spacetime has been assumed to be spatially homogeneous and isotropic \ie a Friedmann-Robertson-Walker spacetime with scale factor $a$.
The evolution of the scale factor is then determined by the Friedmann equations
\begin{equation}
\begin{aligned}
\ddot{a} &= -\frac{4\pi Ga}{3}\langle\rho+3p\rangle\\
\Big(\frac{\dot{a}}{a}\Big)^2 &= \frac{8\pi G}{3}\langle\rho\rangle\\
\end{aligned}
\end{equation}
where the energy density and pressure are given by
\begin{equation} \label{eq:rho-pres}
\begin{aligned}
\rho &= \sum_{i}\Big(\frac{1}{2}\dot{\phi_{i}}^2 + \frac{1}{2a^2}(\nabla\phi_{i})^2\Big) + V(\phi_{i})\\
p &= \sum_{i}\Big(\frac{1}{2}\dot{\phi_{i}}^2 - \frac{1}{6a^2}(\nabla\phi_{i})^2\Big) - V(\phi_{i})\\
\end{aligned}
\end{equation}
and $\langle\rangle$ denotes average over the volume of the lattice. Note that $\nabla$ is the gradient operator with respect to comoving coordinates. 

CUDAEASY uses a staggered leapfrog method to solve these equations of motion. %This was also used in LATTICEEASY \cite{Felder:2000hq}.
This means that the field values and their derivatives are stored at different times and they are advanced in turns:
%The actual leap-frog steps that we use to evolve the fields are
%\begin{equation} \label{eq:evo}
%\begin{aligned}
%& \cdots \\
%\phi_{pr}(t_{pr}) \quad &= \quad \phi_{pr}(t_{pr} - dt_{pr}) + dt_{pr} \phi_{pr}'(t_{pr} - dt_{pr}/2)\\
%\phi_{pr}'(t_{pr} + dt_{pr}/2) \quad &= \quad \phi_{pr}'(t_{pr} - dt_{pr}/2) + dt_{pr} \phi_{pr}''(t_{pr})\\
%\phi_{pr}(t_{pr} + dt_{pr}) \quad &= \quad \phi_{pr}(t_{pr}) + dt_{pr} \phi_{pr}'(t_{pr} + dt_{pr}/2).\\
%& \cdots \\
%\end{aligned}
%\end{equation}
\begin{equation} \label{eq:evo}
\begin{aligned}
& \cdots \\
\phi(t) \quad &= \quad \phi(t - dt) + dt \dot{\phi}(t - dt/2)\\
\dot{\phi}(t + dt/2) \quad &= \quad \dot{\phi}(t - dt/2) + dt \ddot{\phi}(t)\\
\phi(t + dt) \quad &= \quad \phi(t) + dt \dot{\phi}(t + dt/2).\\
& \cdots \\
\end{aligned}
\end{equation}
For second order differential equations this method is however stable only if the corresponding differential equation has no first order time derivative terms, \ie $\ddot{\phi} = \ddot{\phi}(\phi)$. We will therefore move into units that will transform equation (\ref{eq:fe}) into suitable form. This was presented in the LATTICEEASY documentation \cite{Felder:2000} and we will only cite the results. The transformed variables read
\begin{equation}
\phi_{pr}=Aa^{r}\phi, \quad \vec{x}_{pr} = B\vec{x}, \quad dt_{pr} = Ba^{s}dt
\end{equation} 
where the rescaling variables $A$, $B$, $r$ and $s$ can be set freely with the limitation that $s-2r+3=0$ in order to eliminate the $\phi_{pr}'$ term from the equation of motion. The scalar field equations then read in the new variables
\begin{equation} \label{eq:scalar-evo}
\phi_{pr}'' - a^{-2s-2}\nabla^2_{pr}\phi_{pr} - \Big(r(s-r-2)\big(\frac{a'}{a}\big)^2+r\frac{a''}{a}\Big)\phi_{pr}+\frac{\partial V_{pr}}{\partial \phi_{pr}}=0
\end{equation}
where
\begin{equation}
V_{pr} = \frac{A^2}{B^2}a^{-2s+2r}V,
\end{equation}
and $V$ is also expressed in terms of the rescaled scalar fields $\phi_{pr}$.

%There are some helpful guidelines given in the LATTICEEASY document when setting the recaling variables.
There are now a range of different values these scaling parameters can take \cite{Felder:2000} while the limitation $s-2r+3=0$ is met.
For example by setting $A=1/\phi_{0}$ where $\phi_{0}$ is the value of the initially dominant scalar field will make the scalar fields of order of unity initially. If it is possible to identify a dominant potential term and to write it as
\begin{equation}
V = \frac{cpl}{\beta}\phi^{\beta}
\end{equation}
the recaling variables can be set to
\begin{equation}
A=\frac{1}{\phi_{0}}, \quad B = \sqrt{cpl}\phi_{0}^{-1+\beta/2}, \quad r = \frac{6}{2+\beta} \quad \textrm{and} \quad s = 3\frac{2-\beta}{2+\beta}.
\end{equation} 

The scale factor evalution in program units is determined by \cite{Felder:2000}
\begin{equation} \label{eq:scale}
a'' = (-s-2)\frac{a'^2}{a} + \frac{8\pi G}{A^2}a^{-2s-2r-1}\Big\langle \sum_{i} \frac{1}{3}|\nabla_{pr}\phi_{i,pr}|^2+a^{2s+2}V_{pr}\Big\rangle
\end{equation}
which we can now use in the leap frog method to evolve the scale factor. Note that this is done differently in DEFROST where the evolution of $L \equiv 1/H$ is solved first and the scale factor $a$ evolution is solved from $L$. This has the advantage that the computationally complex gradient terms cancel out from the equations of motion. We chose to use the leap frog method mainly for pragmatic reasons: the lattice simulations are usually done with periodic boundary conditions and when calculating the average over the volume of gradient terms we can use the divergence theorem to get
\begin{equation}
\int{|\nabla\phi_{i}|^2 dV} = \int{\nabla\phi_{i}\cdot\nabla\phi_{i} dV} = \int{(\phi_{i}\nabla\phi_{i}) \cdot d\vec{S}}-\int{\phi_{i}\Delta\phi_{i}dV}
\end{equation}
and the surface integral cancels out due to periodic boundary conditions. We can now use this result in the volume averages of the gradient terms in scale factor equation (\ref{eq:scale}) and since the Laplacian is already calculated in the scalar field equations (\ref{eq:scalar-evo}) the computation of the average gradient term actually has practically no computational effect. This turns out to be useful information in the GPU implementation. Note that this is valid only for the average gradient terms. When calculating the energy density and the pressure we still need to calculate the computationally difficult gradient terms.

The initial values of the scalar fields are computed as presented in DEFROST \cite{Frolov:2008hy} \ie the homogeneous scalar field values are given as an input and the random quantum fluctuations are created by the program. The main differences come from the use of CUDA Fast Fourier Transform (CUFFT) instead of FFTW (an acronym for Fastest Fourier Transform in the West) and from the fact that we need to transform the scalar fields into program units that the leap frog uses \ie
\begin{equation}
\begin{aligned}
\phi_{pr} &= Aa^r\phi,\\
\phi_{pr}' & = \frac{A}{B}a^{r-s}\dot{\phi}+r\frac{a'}{a}\phi_{pr}.\\
\end{aligned}
\end{equation}
After the initial values have been set we need to make the field values and their derivatives desynchronized by advacing the the scalar fields $dt_{pr}/2$ steps forward.

\subsection{CUDA implementation}

As mentioned before GPUs are good at doing parallel computations whereas CPUs are much faster at serial calculations.
We have therefore divided the computational labor between the GPU and the CPU based on how parallel different parts of the problem are.
The scalar field equations (\ref{eq:scalar-evo}) are the hardest part computationally but they can be parallelized easily since similar leap frog steps are taken at every point of the lattice with the same equations of motion. The evolution of the scale factor on the other hand is easy to evolve serially once the averages of the gradient and potential terms have been calculated. We will therefore solve the evolution of the scalar fields on the GPU and leave the scale factor evolution to the CPU.

We will start by discretizing the equations of motion (\ref{eq:scalar-evo}) and (\ref{eq:scale}) in a cubic $n^3$ lattice in comoving coordinates with periodic boundary conditions. Since the computations on the GPU are done in single precision we have taken some additional steps to increase the numerical accuracy of the method. Compared to the naïve discretization of the Laplacian used by LATTICEEASY we employ a better one presented by Patra and Karttunen in \cite{Patra:2005} that is second-order accurate and fourth-order isotropic. This more complex discretization uses all 26 neighbours of a 3x3 cube compared to the six closest neighbours the naïve discretization uses. This same stencil is also used in DEFROST \cite{Frolov:2008hy}.

%The Laplacian in LATTICEEASY is discretized with the naïve method that uses six neighbours of a point. As pointed out in DEFROST \cite{Frolov:2008hy} this is not the best choice since it is not isotropic and will therefore lead to anisotropic errors. Much better discretizations are presented by Patra and Karttunen in \cite{Patra:2005}. We will use a stencil, where all 26 neighbours of a 3x3 cube around a point are needed. This scheme is second-order accurate and fourth-order isotropic and is also used by DEFROST.

\begin{figure}[t]
\subfigure[]{\label{fig1a}\includegraphics*[width=0.45\columnwidth]{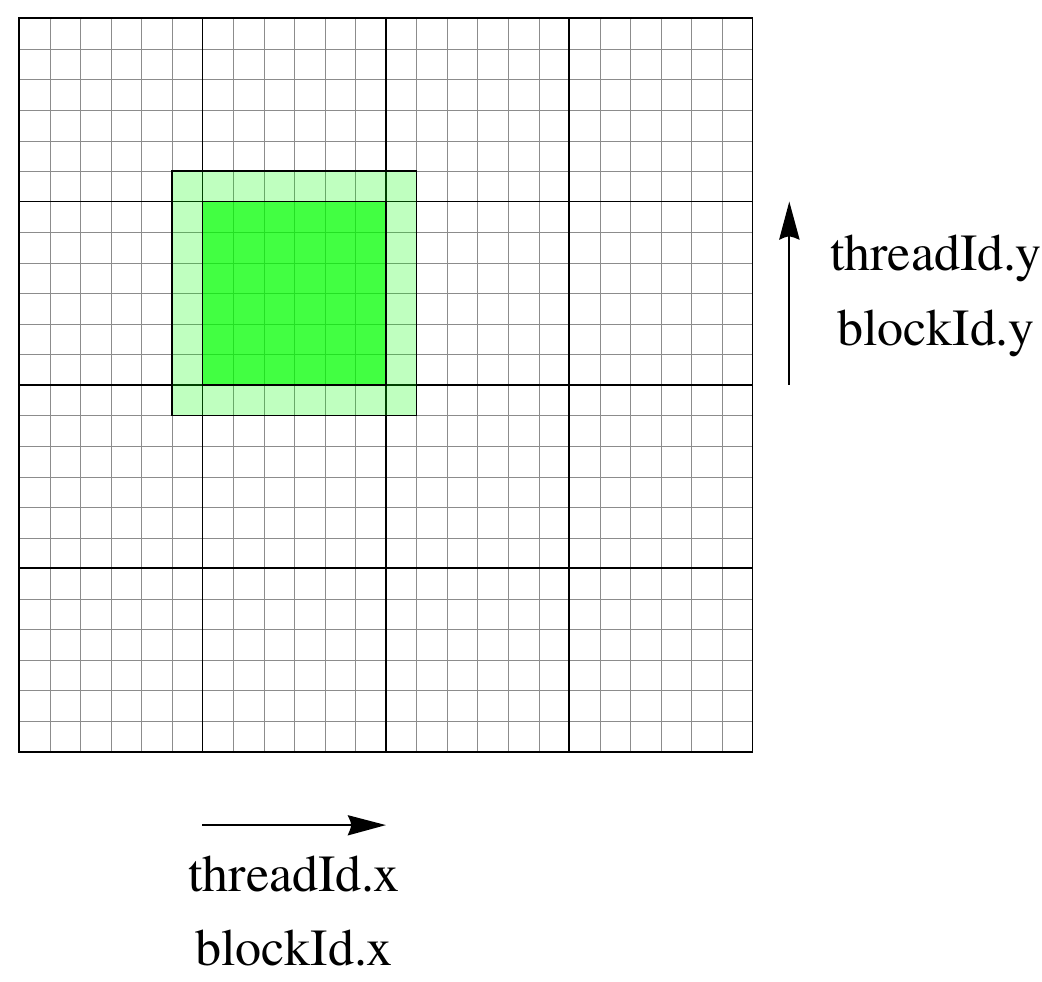}}
\quad
\subfigure[]{\label{fig1b}\includegraphics[width=0.45\columnwidth]{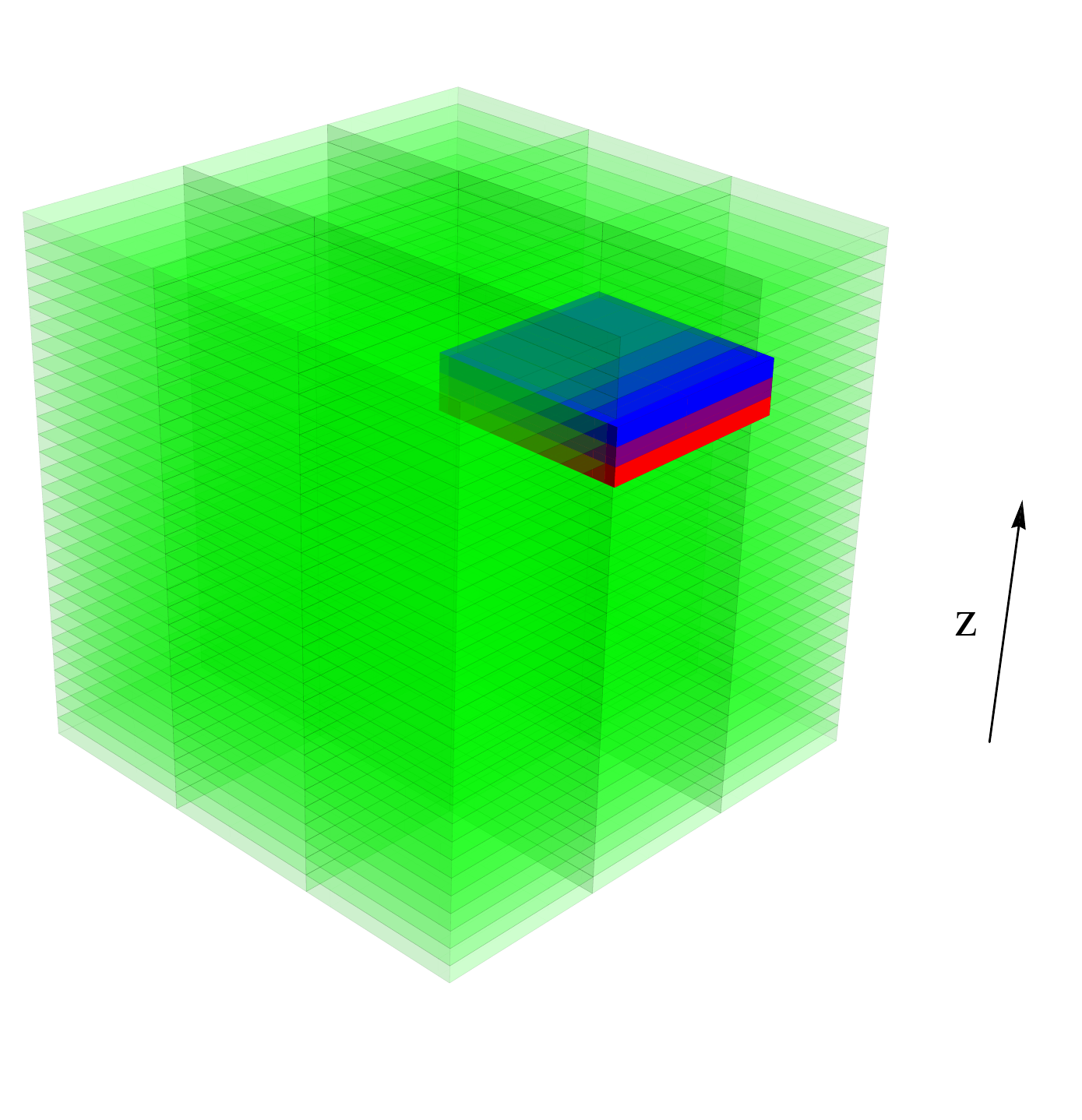}}
\caption{(a) Illustration of the CUDA grid with 16 thread blocks and 64 threads per block. The actual compute grid used in the program is much larger. The computations are done by the inner threads (green) of a block whereas the outer threads (light green) are used only to read data. Note that the thread blocks intersect each other.
(b) Illustration of how data in the shared memory is updated. Thread blocks advance in $z$-direction and load data into the blue slice (shared\_up) from the global memory. The previous values of blue slice are stored into the purple one (shared\_middle) and the values from the middle slice are stored into the red block (shared\_down). This way only a fraction of the memory bandwidth is needed.}
\end{figure}

CUDA implementation of 3D stencils was presented in detail in \cite{Micikevicius:2009}. We will follow a similar method that uses shared memory to calculate the Laplacians. Because of the limited amount of shared memory per SM (16 KB) the 3 dimensional lattice has to be divided into smaller pieces that fit into the shared memory. To accomplish this we have sliced the lattice along the $z$-axis into smaller tiles and advance these tiles in $z$-direction. The computations within these tiles are done by thread blocks. This means that a single thread of a thread block will advance all the scalar fields of a column with constant $x$- and $y$-coordinates. Since the outer threads of a thread block need the values of scalar fields that are in a different block and since different blocks can only communicate through global memory the role of the outer threads of a thread block is only to load data into shared memory and the scalar field computations are done by the inner threads (see figures \ref{fig1a} and \ref{fig1b} for illustration). This implementation unfortunately means that the reads are not fully coalesced and therefore additional speedups might be possible. One way would be to use texture memory to store the values of the scalar fields. This has not been implemented yet.

An alternatice way to update the slices would be to do the calculations with all of the threads of a block and make some threads load more than one value into shared memory. We however noticed that this leads to thread branching within warps and numerous integer computations that slow the computations considerably and negate the speedup the additional threads would give. The overlapping of the threads in the current method however means that the thread indexes are not linear in the CUDA grid and extra care has to be taken to keep the globar memory reads consistent. We have done this with a device function that also takes into account the periodic boundary conditions. 

In order to reduce global memory usage every block uses three layers of shared memory: \textsl{shared\_down}, \textsl{shared\_middle} and \textsl{shared\_up} respectively. The tiles advance in $z$-direction and load new data into the \textsl{shared\_up} memory from global memory. Before doing this update the previous values of \textsl{shared\_middle} are stored into \textsl{shared\_down} and the values from \textsl{shared\_up} are stored in the middle memory. This updating method will only need a fraction of the memory bandwith compared to a naïve method where every thread would do 27 global memory loads for one calculation. Current implementation with 256 threads in a thread block needs roughly only $\approx 1/21$th of this banwidth and therefore makes the kernel much faster. The computation starts from the bottom of the lattice \ie $z=0$ and due to the periodicity of the lattice the initial values of \textsl{shared\_down} are loaded from the top of the lattice. See fig \ref{fig1b} for an illustration of the update method.

After the different shared memory slices have been updated the Laplacian can be calculated with the discretization mentioned before. Since the coefficients used in this dicretization are the same at every point of the lattice we have stored these in the constant memory of the GPU which is meant to store read-only data for the threads. This way we can lower the register need of a kernel and make the program faster.

Once the Laplacian is calculated we can proceed to the actual leap frog step \ie equations (\ref{eq:evo}). The other terms beside the Laplacian depend only on the local values of the scalar fields and can therefore be stored in registers. Those terms in equation (\ref{eq:scalar-evo}) that either depend on the scale factor or are constants such as any factors in the potential derivatives are stored in the constant memory since these are constant throughout the lattice during a leap-frog step. This way we can save some of the registers but also avoid any divisions on the GPU which are much slower than additions or multiplications. Once the leap-frog steps have been taken the new values of the scalar field and its derivative are written into global memory and are used in the next step.

Before advancing a thread block in $z$-direction we will do some additional calculations needed in the scale factor evolution: namely we need to calculate the averages of the squared gradient and the potential term in equation (\ref{eq:scale}). As explained previously the gradient term can be written in terms of the Laplacian and since this has been already calculated the gradient term is trivial to implement in the kernel. We simply make a new variable (called $gpe$ for gradient and potential energy in the program) and subtract the values of $\phi_{pr}\Delta_{pr}\phi_{pr}$ from it along the column the thread is advancing in $z$-direction. The potential terms are done similarly but they are written only for the last scalar field being evolved. We will employ additionally the Kahan summation \cite{Kahan:1965} when calculating these sums in order to keep the numerical errors as small as possible.

Once the calculations needed in the averages are done the thread block advances in $z$-direction and loads new values into shared memory, calculates the Laplacian, advances the scalar fields and increments the $gpe$ variable and advances again. Once the thread blocks reach the top of the lattice the evolution step finishes for the lattice and writes the values of $gpe$ into global memory. This way the global memory bandwidth needed is reduced significantly compared to a method where the increment variables would be written at every point of the lattice since the values are now written only once by every thread.

After finishing the evolution kernel for one scalar field the same calculations need to be done for the other fields as well. All of the fields could be in theory advanced with one kernel call but this would quickly lead to an extremely long kernel function suitable only for a very spesific model. Current implementation uses two fields but this can be easily increased for different more complex models. The scale factor evolution on the CPU is done once every scalar field has been evolved.

We have also written a more complex kernel function that is used to calculate additionally the energy density and pressure of the scalar fields \ie equations (\ref{eq:rho-pres}). The discretization of these was presented in \cite{Frolov:2008hy} and the implementation in CUDA is similar to the Laplacian method explained previously. Since this kernel does more calculations and uses more registers the main evolution is done with the simpler and faster kernel. How often the energy density and pressure need to be calculated is left to the user. These quantities are written by the program in SILO format into seperate files that can be easily visualized with the VISIT program provided by LLNL (see figures \ref{fig3a}-\ref{fig3d}).

\section{Numerical results}

In order to test the program and verify the output we have solved the evolution of scalar fields in the simple chaotic inflation model
which was studied in \cite{Podolsky:2005bw} with LATTICEEASY and in \cite{Frolov:2008hy} with DEFROST.
The potential in this case is
\begin{equation}
V(\phi,\psi) = \frac{1}{2}m^2\phi^2 + \frac{1}{2}g^2\phi^2\psi^2,
\end{equation}
where $\phi$ is the inflaton and $\psi$ the decay product.
We will use units similar to DEFROST \ie we will set the reduced Planck mass $m_{\textrm{pl}}=(8\pi G)^{-1/2}$ equal to one in the computations. The initial values have been set to coincide with the values given in \cite{Podolsky:2005bw,Frolov:2008hy}. This means that homogenous value of the inflaton is set to $\phi \simeq 1.009343$ and the decay field $\psi = 0$, the mass of the inflaton equals $5 \cdot 10^6 m_{\textrm{pl}}$, the Hubble parameter $H \simeq 0.50467m$, and the value of the coupling constant $g=100$. We have evolved the system with two different time steps ($dt = 2^{-9}/m$ and $dt = 2^{-10}/m$) and lattice sizes  ($128^3$ and $256^3$ respectively). The rescaling parameters have been chosen to be
\begin{equation}
A=\frac{1}{\phi_{0}}, \quad B = \sqrt{m}, \quad r = \frac{3}{2} \quad \textrm{and} \quad s = 0,
\end{equation} 
which means that $dt_{pr} = dt$. We have evolved this system with $2^{18}$ time steps which corresponds to evolution times of $t = 512/m$ and $t = 256/m$ depending on the time step and lattice size. The larger lattice corresponds to the one used in DEFROST.

The CUDA kernels are run on a NVIDIA GTX 260 SP 216 with 896 MB of global memory which can be bought at the time of writing for $\sim 200$ \$ from a well equipped computer store. The host computations are run on a quadcore AMD Phenom processor with 4 GB of host memory. We chose to use thread blocks that contain 324 threads of which computations are done by 256 threads. This configuration was found to be the fastest of 64, 128 or 256 threads per block. The CUDA grid size is based on the size of lattice. For the $256^3$ case we chose to use $256 (=16^2)$ thread blocks.

As noted previously the computations are done with two kernels depending on if the energy density and pressure need to be calculated. The simpler of these does most of the calculations whereas the other kernel is cuttently called once in 512 steps. Because of the complicated gradient functions the kernels are under heavy register pressure. The main evolution kernel for example needs 24 registers per thread. Despite our best efforts we haven't yet been succesful in lowering this register pressure. If future generations of GPUs have more registers we expect more speedups from the program.

\begin{figure}[h]
\subfigure[]{\label{fig2a}\includegraphics*[width=0.45\columnwidth]{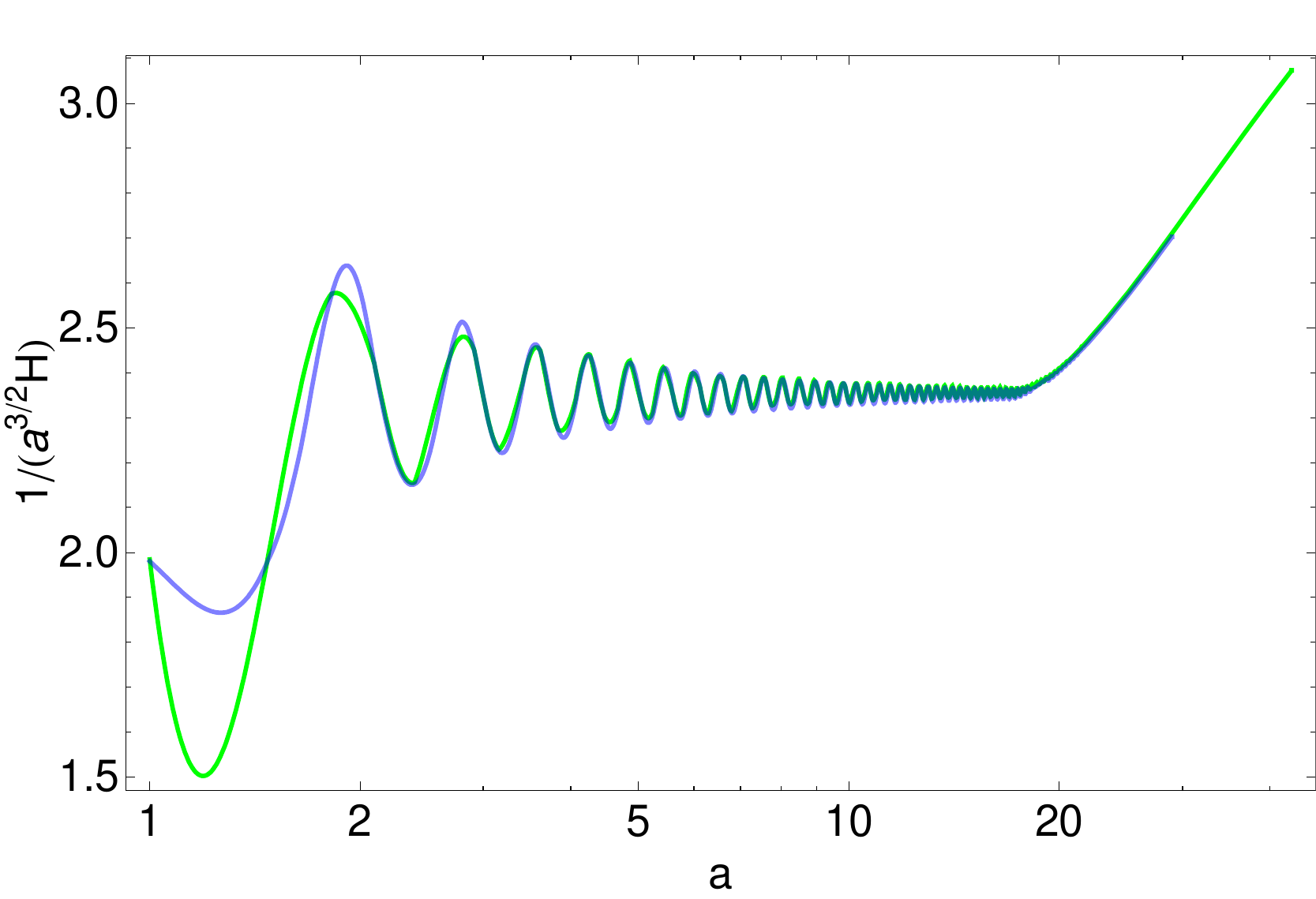}}
\quad
\subfigure[]{\label{fig2b}\includegraphics[width=0.45\columnwidth]{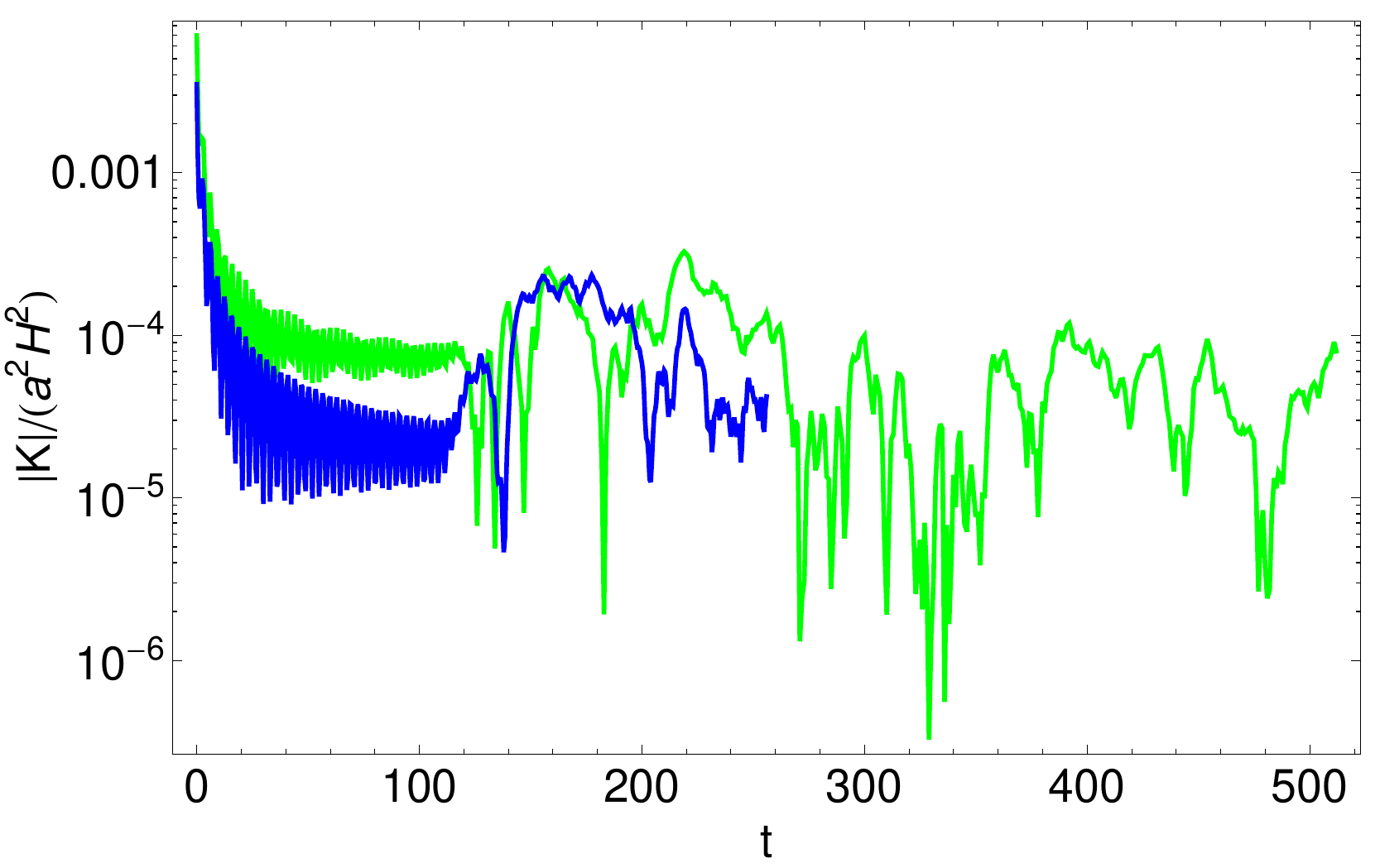}}
\quad
\subfigure[]{\label{fig2c}\includegraphics[width=0.45\columnwidth]{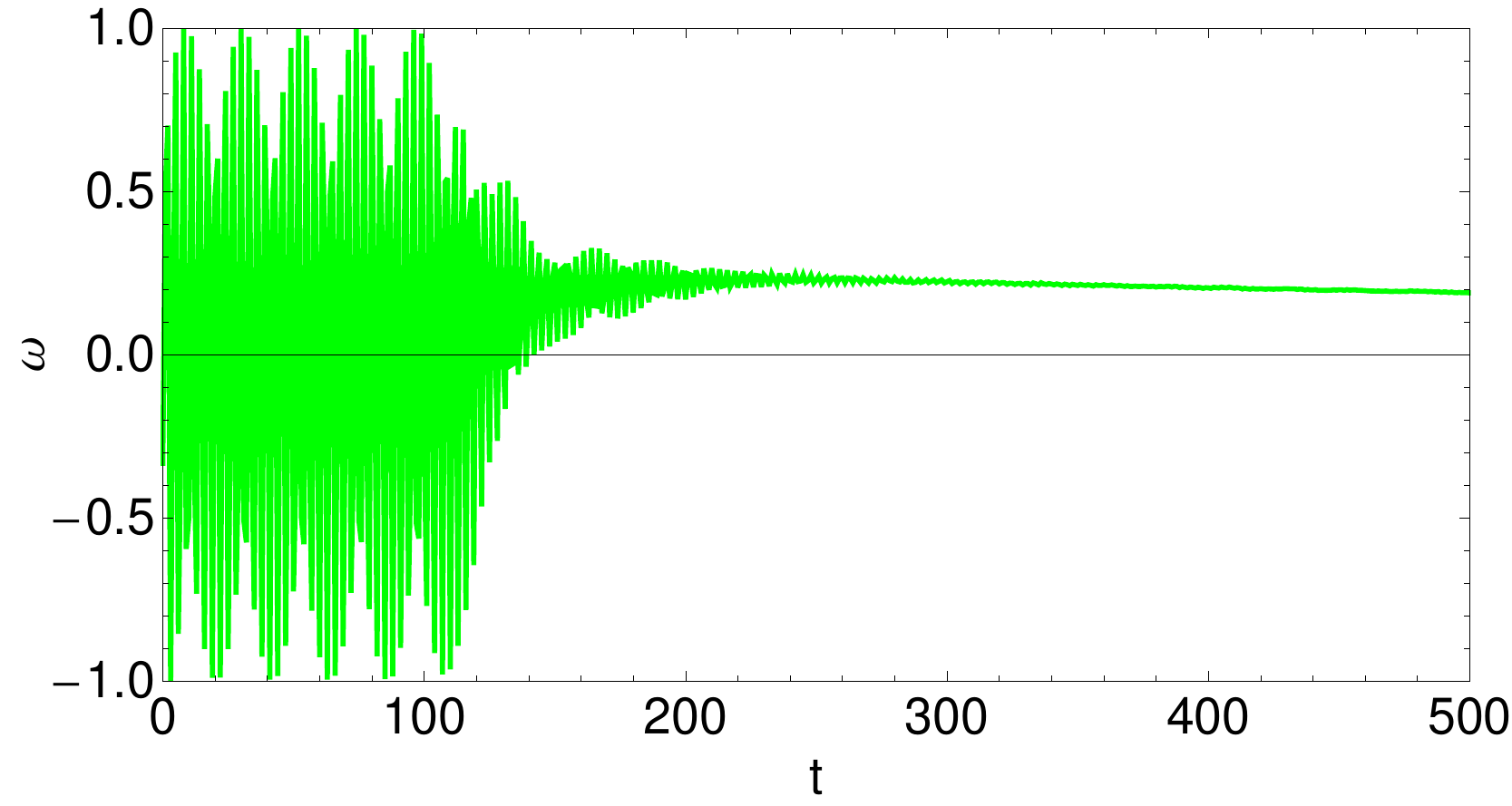}}
\quad
\subfigure[]{\label{fig2d}\includegraphics[width=0.45\columnwidth]{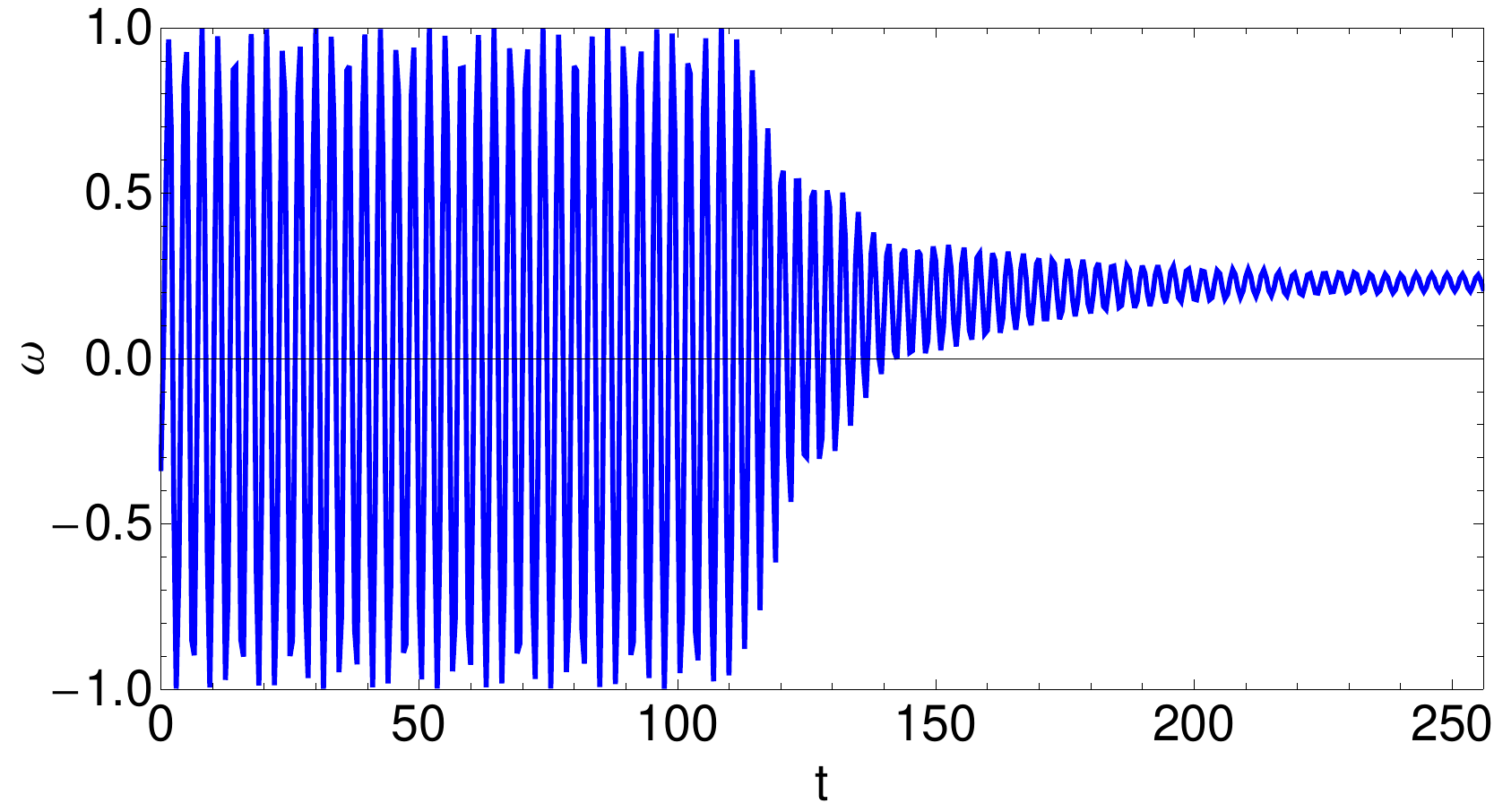}}
\caption{(a) The evolution $1/(a^{3/2}H)$ as a function of the scale parameter in $128^3$ (green) and $256^3$ (blue) lattices.
(b) The absolute value of the relative residual curvature $\frac{K}{a^2H^2} = \Big(\frac{8\pi G \langle \rho \rangle }{3 H^2} - 1\Big)$ in the $128^3$ (green) and $256^3$ (blue) lattices. Once the parametric resonance starts the errors increase slightly but these are generally very small considering that the evolution is calculated in single precision.
(c), (d)
The evolution of the effective equation of state $w = \langle P \rangle / \langle \rho \rangle$ in the lattice of (c) $128^3$ and (d) $256^3$ elements.}
\end{figure}

\begin{figure}[h]
\subfigure[]{\label{fig3a}\includegraphics*[width=0.45\columnwidth]{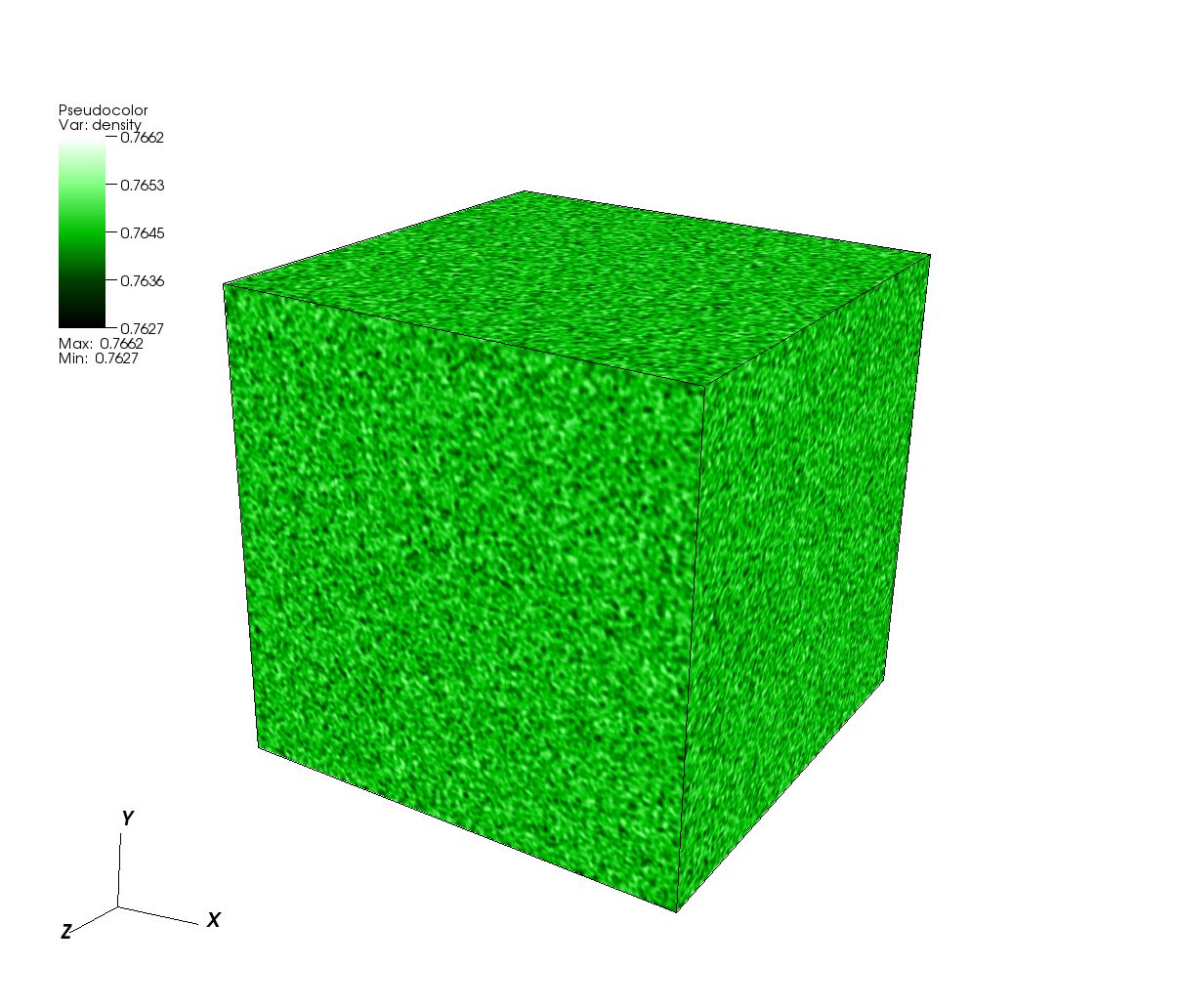}}
\quad
\subfigure[]{\label{fig3b}\includegraphics[width=0.45\columnwidth]{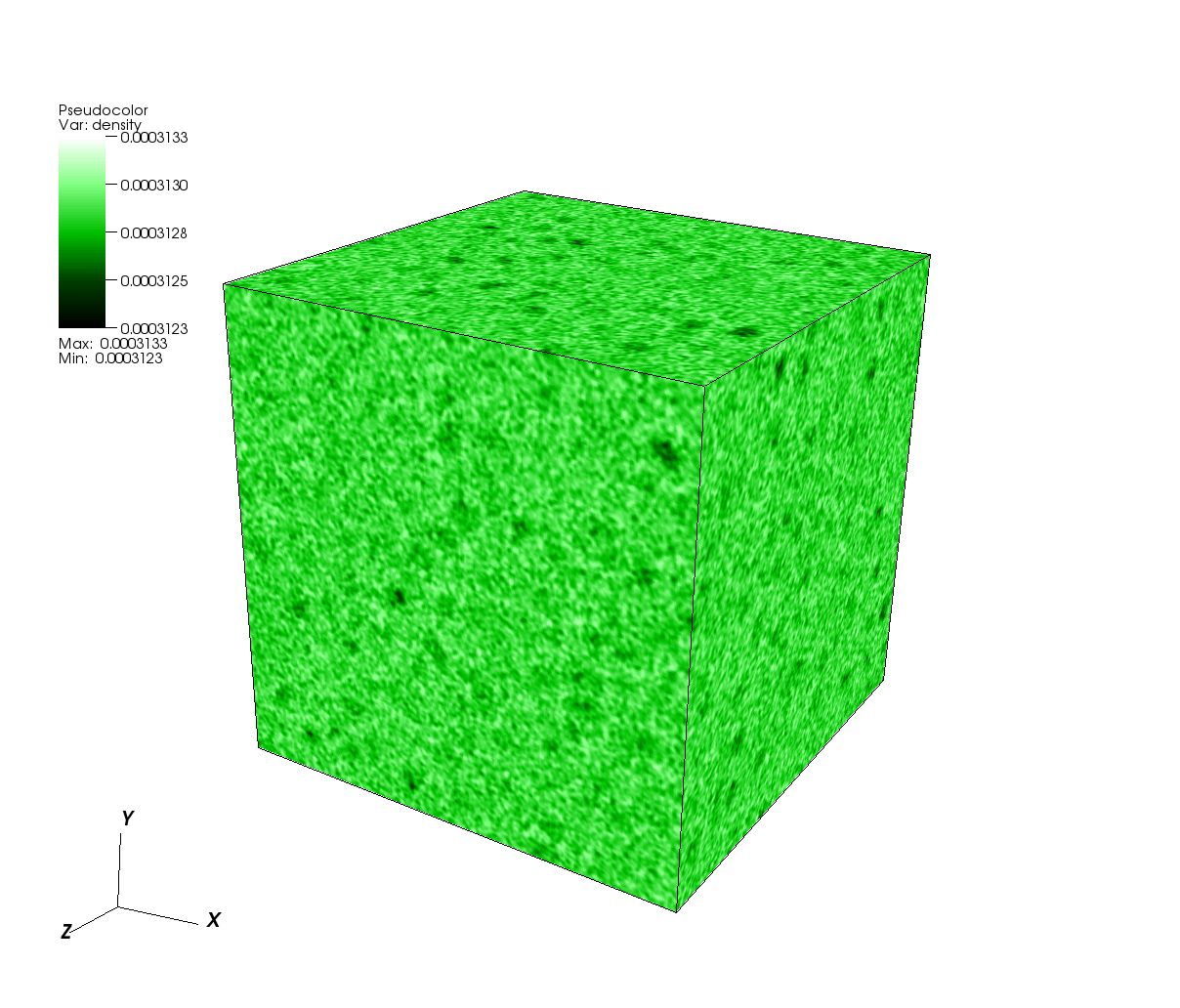}}
\quad
\subfigure[]{\label{fig3c}\includegraphics[width=0.45\columnwidth]{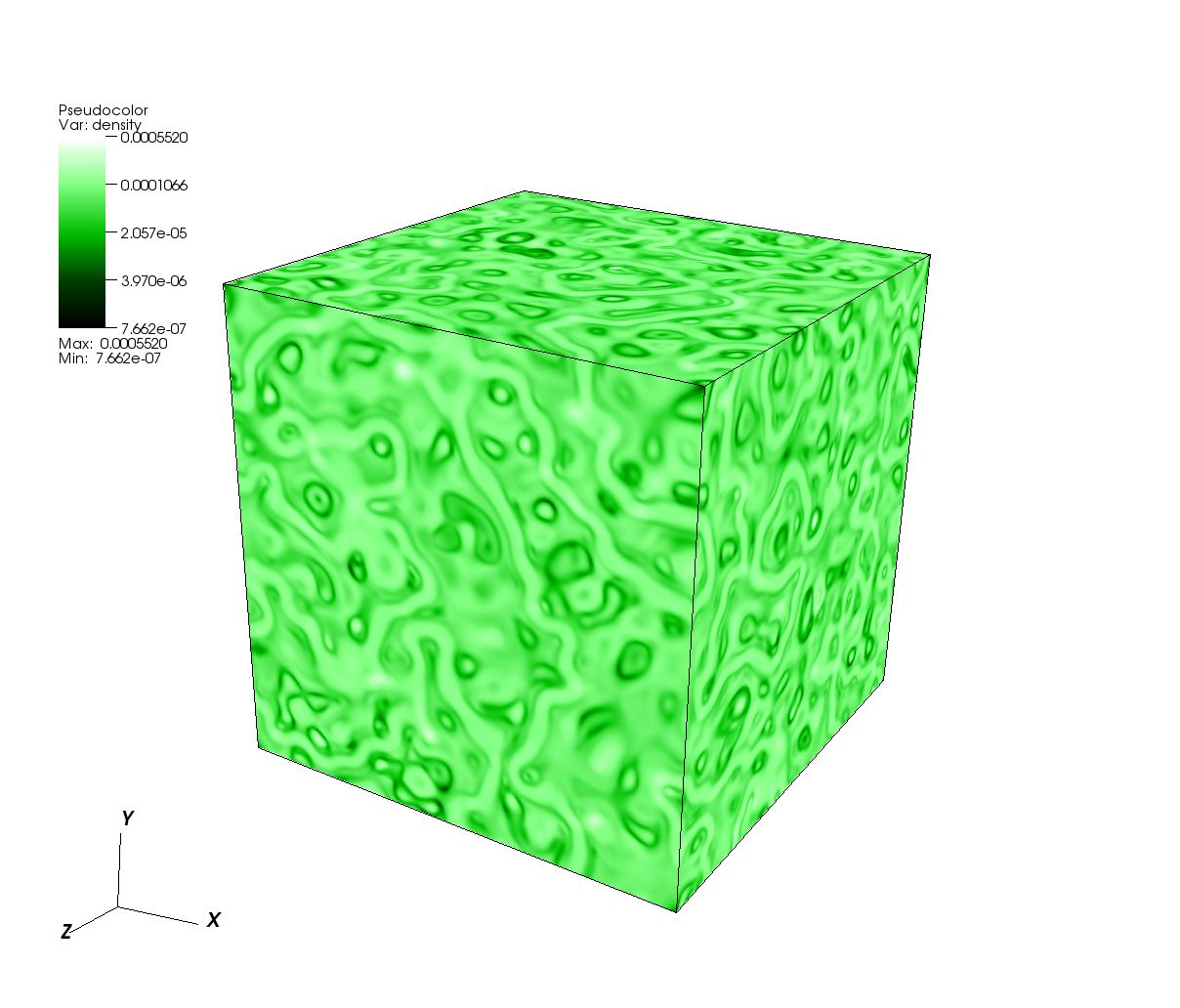}}
\quad
\subfigure[]{\label{fig3d}\includegraphics[width=0.45\columnwidth]{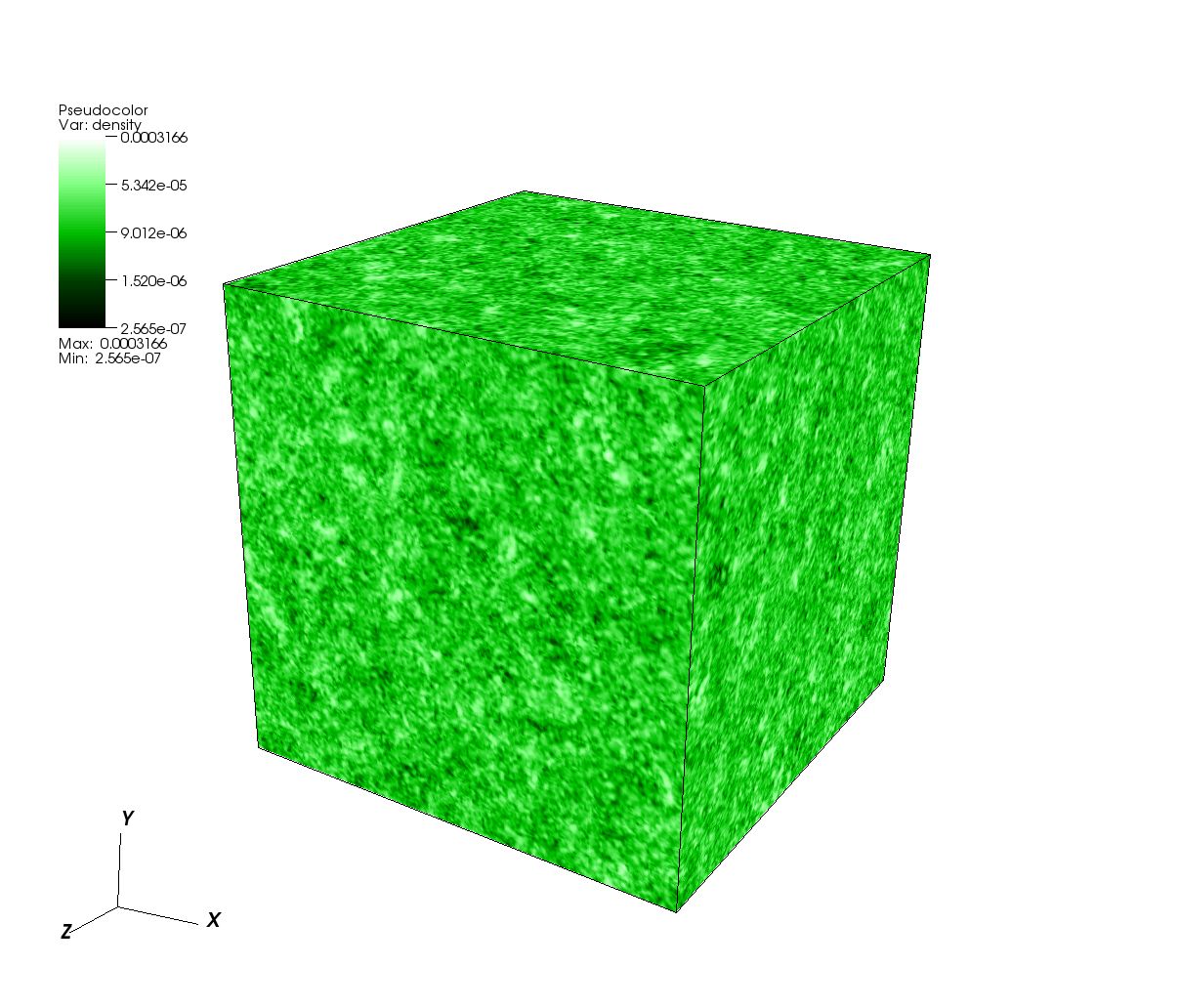}}
\caption{The evolution of energy density during simulation. Plots show logarithm of energy density at (a) $t=0$, (b) $t=64/m$, (c) $t=128/m$ and (d) $t=256/m$. The start of parametric resonance can be clearly seen in figure \ref{fig3c}. After the resonance energy densities dilute to the state shown at \ref{fig3d}.}
\end{figure}

The numerical results are shown in figures \ref{fig2a}-\ref{fig3d}. Fig. \ref{fig2a} shows the evolution $1/(a^{3/2}H)$ as a function of the scale parameter with two different lattice sizes: the evolution is plotted in green for $128^3$ lattice and in blue for $256^3$ lattice. Initially the systems evolve differently but as time progresses the the same evolution is restored. The difference is mainly due to different initializations. These figures show that the system produces the same evolution as the DEFROST program. In fig. \ref{fig2b} we have plotted the absolute value of the relative residual curvature
\begin{equation}
\frac{K}{a^2 H^2} = \Big(\frac{8\pi G \langle \rho \rangle }{3 H^2} - 1\Big)
\end{equation}
as a function of time in the different lattices which we use as a proxy for numerical errors. The numerical errors clearly dilute as time progresses altough the parametric resonance silghtly increases it. Considering that the field equations are solved in single precision residual errors this small can be considered exceptional. Current version of the code could be easily made to run also in double precision but this would drop the performance to one eight of the current throughput due to the number of double precision units on the chip. The size of global memory would also limit the size of the lattice to be smaller in double precision.

Besides the system evolution we have also plotted 3D visualizations of the evolution of the energy density in the $256^3$ lattice in figures \ref{fig3a}-\ref{fig3d}. These figures were made with LLNL's Visit visualization program. The plots show the logarithm of energy density at $t=0, t=64/m, t=128/m \textrm{ and } t=256/m$ respectively. The start of parametric resonance can be clearly seen in figure \ref{fig3c}.

The total computation time per one step is roughly 0.022 seconds which corresponds to a total running time of one hour and 36 minutes for one simulation with $256^3$ elements. The smaller lattice $128^3$ with $2^{18}$ steps is solved in less than 900 seconds which corresponds to less than 0.0034 seconds per one time step. Compared to the DEFROST program this is an order of magnitude faster (DEFROST reports a computation time of 0.32 seconds per time step on an Intel Xeon computer for the $256^3$ lattice). Compared to LATTICEEASY this speedup increases by a factor of four \cite{Frolov:2008hy}. What this means is that simulations that used to last roughly one day can now be computed in hours and in the future this speedup is expected to increase even more due to software improvements and due to the more rapid increase of computational power of GPUs compared to CPUs \cite{Belleman:2007kv}.

\section{Discussion and conclusions}

We have presented a GPGPU implementation of a program that solves the evolution of interacting scalar fields in an expanding space. The main implementation has been done in the spirit of LATTICEEASY which means that users familiar with aforementioned program should find it relatively easy to start using CUDAEASY. We have implemented most of the improvements that were first presented in DEFROST including making the initializations of the program consistent and using more advanced discretizations of differential operators. Because of these and other improvements the program achieves roughly same precision as DEFROST while using single precision in the GPU calculations.

We would like to note that we have not (at least yet) implemented all of the functionalities of LATTICEEASY and DEFROST: namely the current version of CUDAEASY doesn't yet produce any quantative data about the spectrum but this will be rather trivial to implement with the CUDA FFT as a post evolution procedure. This is however left to future versions of the program.

Other improwements might include a multi-GPU version of CUDAEASY that could be used to do simulations in much larger lattices. Periodic boundary conditions and the initialization of the lattice might pose some problems but these questions are beyond the scope of this paper. Graphics processing units also make it possible to visualize the evolution of the system in real time: one example are the N-body simulations presented in \cite{Belleman:2007kv}. In CUDAEASY this could be applied to smaller lattices that are simulated in a few minutes. For larger ($\sim 256^3$) lattices the real time visualization might however have a performance hindering effect.

Another interesting venture would be the porting of CUDAEASY into OpenCL. At the time of writing there are however some obstacles before this can be done though. Despite our best efforst we haven't yet found a solid FFT implementation on OpenCL. Once this situation improves CUDAEASY should be relatively easy to port to OpenCL language and therefore harness the computational power of AMD GPUs also into preheating calculations.

Since the aim of this paper was to present a GPU implementation of LATTICEEASY and DEFROST we have not done any original research in cosmology. This is the aim of a future paper where we will study non-perturbative curvaton decay with CUDAEASY. This has been previously studied in \cite{Chambers:2009ki} where $\Delta N$ formalism was used to calculate the non-gaussianity \cite{Enqvist:2004ey, Chambers:2007se,Chambers:2008gu} generated by the resonant curvaton decay. We believe that with CUDAEASY the needed simulations can now be run much faster without sacrificing the accuracy of the results.

Besides solving preheating evolution CUDAEASY can also be used after slight modifications to study the evolution of Q-balls \cite{Multamaki:2002hv}, gravity waves, phase transitions, black hole production and many more interesting phenomena in early universe \cite{Felder:2008}.
%Lately computational modelling has increased drastically in science in general. Since the number of supercomputers is quite limited there is a market for so called high performance computing. GPUs can deliver this at a fraction of price and at desktop level and thereby leverage the computational horsepower of ordinary PCs considerably.
With the computational power of GPUs many of these problems can now be studied much faster than previously even on a common desktop computer without losing accuracy. We hope that CUDAEASY is able to achieve this ambitious goal and be useful to the interested user.

\subsection*{Acknowledgments}
JS is supported by the Academy of Finland project no. 8111953. The author would like to thank Iiro Vilja and Joonatan Palmu for valuable discussions when writing the CUDAEASY program.

%\newpage

%%%%%%%%%%%%%%%%%%%%%%%%%%%

\end{document}